\documentclass{article}
\usepackage[preprint]{log_2022}

\usepackage{booktabs}           
\usepackage{multirow}           
\usepackage{amsfonts}           
\usepackage{graphicx}           
\usepackage{duckuments}         
\usepackage{wrapfig,lipsum,booktabs}

\usepackage[
     backend=biber,
     style=numeric-comp,
      backref=true,
     natbib=true]{biblatex}
\title[A1]{A Potts model approach to unsupervised graph clustering with Graph Neural Networks}

\author[A2]{%
Co Tran\\
\institute{SailPoint Technologies, Austin}\\
\email{cotran2\@utexas.edu}\And
Mo Badawy\\
\institute{SailPoint Technologies, Austin}\\
\email{m.cohomology\@gmail.com} \And
Tyler McDonnell\\
\institute{SailPoint Technologies, Austin}\\
\email{tyler.mcdonnell\@gmail.com}
}
\newcommand\normx[1]{\left\Vert#1\right\Vert}

\begin{document}

\maketitle

\begin{abstract}
Numerous approaches have been explored for graph clustering, including those which optimize a global criteria such as modularity. More recently, Graph Neural Networks (GNNs), which have produced state-of-the-art results in graph analysis tasks such as node classification and link prediction, have been applied for unsupervised graph clustering using these modularity-based metrics. Modularity, though robust for many practical applications, suffers from the resolution limit problem, in which optimization may fail to identify clusters smaller than a certain scale that is dependent on properties of the network. In this paper, we propose a new GNN framework which draws from the Potts model in physics to overcome this limitation. Experiments on a variety of real world datasets show that this model achieves state-of-the-art clustering results.
\end{abstract}
\section{Introduction}
Graph clustering, also referred to as community detection, has been utilized in a multitude of practical applications in biology \cite{gc_bio}, social networks \cite{gc_soc}, neuroscience \cite{gc_neurosc}. According to \cite{zhao2012consistency}, generally speaking, there are three identifiable themes to graph clustering, also sometimes referred to as community detection. 

One approach utilizes classical, greedy algorithms, e.g. hierarchical clustering, see \cite{newman2004detecting}. Second class of methods incorporates a properly selected global function or criteria, the communities are then identified via an optimization process of such function over the set of all possible partitions of the network graph, e.g. graph cuts \cite{graphcuts1}, \cite{graphcuts2}, spectral clustering \cite{spectral}, \cite{mincut}, modularity-based methods \cite{newman2006modularity}, \cite{spectral_modularity_optimization}, etc. The third approach makes use of probabilistic modeling of the network, where a probabilistic model is learned or approximated to maximize the likelihood for specific graph labeling configurations that yield the desirable clustering results. An example of the this would be stochastic block methods \cite{stochblock}, along with the degree-corrected version \cite{stochblock_degcorrected}, and random cluster models \cite{random_cluster}. 

Modularity-based methods, e.g. Louvain, see \cite{louvain}, have gained popularity due to relative simplicity of the algorithm, ease of scalability for large graphs, and robustness of results. The Newman-Girvan modularity is easy to define for a given graph labeling, measuring how likely the labeling would have occurred by chance. Identifying an optimal labeling (node partition) that maximizes modularity typically yields good clustering results. Modularity approaches suffer from some drawbacks, most notably the resolution limit issue, see \cite{resolutionlimit}. The resolution limit issue makes it harder harder to identify relatively small clusters in larger graphs.

In this paper, we present two main contributions. The first is achieving new SoTA results compared to \cite{dmon}. Our approach utilizes a GNN network that is used to optimize a criteria derived from a Potts model of the given graph. Potts models have been already used as random graph models for graph clustering purposes, see \cite{potts_temp}. Moreover, previous work, \cite{potts_reslimfree}, shows that they are not prone to the resolution-limit issues that affect the traditional Girvan-Newman modularity-based approaches. 
The second contribution is a by-product of utilizing the Potts model approach in that it allows us to manipulate, and potentially optimize, certain model parameters, e.g. temperature as in \cite{potts_temp} to control the level of granularity or resolution of the resulting clustering. Such parameters lend themselves in a natural way to the clustering problem enabling the researcher to identify "reasonable" values and thus allows a completely unsupervised solution to the clustering problem. This can be useful in many applications where knowing the number of clusters, a priori, is not feasible.

This paper is structured as follows, in section one ...

\section{Motivation}
In recent years, graph neural networks have been utilized to perform various tasks on graph data structures, e.g. node classification, link/edge prediction. The Message passing framework for GNNs, see for example \cite{gnns_msgpassng}, has been successful in generalizing earlier architectures such as Convolutional and Attention GNNs. This message-passing paradigm allows the node and edge properties to be locally pooled within the localized node neighborhoods, thus enabling the GNN to learn representations of the local and higher-order structure of the graph. Intuitively, one would think that pooling could be enough to generate adequate representations to solve graph clustering problems, but that has been shown to be false in the recent paper \cite{dmon}, where the DMoN GNN was introduced as a (partially) unsupervised approach to solve graph clustering problems, where the number of total clusters is known. The DMoN approach has been shown to achieve SoTA results on standardized graph clustering datasets where the ground truth is known or easily inferred.

Our work in this paper has been motivated so as to contribute two improvements to the previous results achieved in \cite{dmon}. More specifically:
\begin{itemize}
    \item The loss function utilized in \cite{dmon} was derived from a Girvan-Newman modularity criteria. This could potentially result in issues regarding modularity maximization, e.g. resolution-limit problems (see \cite{resolutionlimit}), where relatively smaller clusters could become harder to identify via a modularity-optimization approach. 
    
    \item The unsupervised approach proposed in \cite{dmon} still requires the knowledge of the number of clusters a priori. This can prove to be a challenge in certain practical applications where there is lack of problem-intrinsic or heuristic knowledge to hint towards a reasonable level of cluster granularity.
\end{itemize}

Our research into other possible approaches led us to \cite{RB} where the Potts model was discussed. In \cite{RB} the authors show that the Girvan-Newman modularity occurs naturally as a global criteria corresponding to a particular intuitive choice of a Hamiltonian (energy) function of the system (node labels/configurations). It turns out that minimizing the Hamiltonian function (which is equivalent to maximizing the corresponding global criteria) over the set of all node labels (configurations) of the graph yields optimal clustering results. 

The Potts model has been used in Physics (Statistical mechanics) for a while, see \cite{potts1952}, \cite{potts1990}. It is a spin glass model that generalizes the Ising model which has been used to model ferromagnetic material, where molecules of the magnetic material are modeled as nodes with 2 possible spin values $\{-1,1\}$. The Potts model generalizes that to a finite number, $q \geq 2$, of spin states. The Potts model has been generalized further by K. Fortuin and P. Kasteleyn in 1969 when they introduced the random cluster model, see \cite{random_cluster} for more history and details. 
The Potts model has been studied for its own peculiar and interesting dynamics, for instance the abrupt phase transitions that are observed for large values of $q$, contrasting the smooth transitions exhibited by the Ising model.

The Potts model can be used, as other similar energy based approaches, to identify optimal graph clustering, see \cite{potts_temp}. Furthermore, and what makes the Potts model more appealing, \cite{potts_reslimfree} shows that the absolute Potts model does not suffer from the dreaded resolution-limit issue that is associated with the Girvan-Newman modularity approaches. 

Finally, and according to experiments conducted by the authors of \cite{potts_temp}, which showed that by tweaking the temperature parameter of the Potts model, one can control the level of granularity of the clustering results and achieve accurate approximations of the true labels. This provides the researcher with a natural/intrinsic tool to help identify the optimal number of cluster for a given application. Further work could explore potential avenues to identify optimal values for such parameters. 

\section{Preliminaries}
First, we lay out the notation for the graph clustering problem. The graph is represented as a tuple of vertices and edges, $G = (V,E)$ and number of nodes, edges being $n = |V|, m = |E|$ respectively. The ajacency is a square matrix $\mathbf{A} \in \mathbb{R}^{n\times n}$ with $\mathbf{A}_{ij} = 1$ if there is a link between node $i,j$ and $0$ otherwise. The node features matrix is described as $\mathbf{X} \in \mathbb{R}^{n\times l}$ with $l$ be the number of features.

\section{Potts Model}
Following the works of FK {\bf{[ref]}}, random cluster models were established as a generalization of percolation, Ising, and Potts models, all of which have been introduced earlier. The Ising model was introduced in statistical mechanics to model ferromagnetic materials, featuring two spin states per node. The Potts model generalizes that to a number of $q$ spin states per node. These models were utilized to model different materials, and studied further for their own interesting characteristics, e.g. the abrupt phase transitions exhibited by the Potts model for large values of $q$ contrasting the smooth transitions of the Ising model.

The Hamiltonian, to be minimized, for the Potts model is defined by penalizing two connected nodes if they do not belong to the same cluster. More precisely, see {\bf{[ref]}}:
$$H(\{z_{ki}\}) = \frac{1}{2} \sum_{i=1}^{n} \sum_{k=1}^{q} \sum_{j=1}^{n} z_{ki}(1 - z_{kj} ) k_{ij} \alpha_{ij} = \sum_{(i,j) \text{ neighbors}} k_{ij} (1 - \delta_{ij}),$$
 
where the $\{z_{ki}\}$ represents a given label assignment, $z_{ki} = 1$ if
the $i$-th node belongs to the $k$-th cluster, and zero otherwise, $\delta_{ij} =
\sum_k z_{ki} z_{kj}$. $k_{ij} = k(x_i, x_j) = k(x_i - x_j)$, a kernel function, and $\alpha_{ij} = 1$ if $i != j$ and $i$-th and $j$-th nodes are connected, and zero, otherwise.

\paragraph{}
{\it{Equivalently}}, see {\bf{[ref]}}, let $q$ be an integer satisfying $q \ge 2$, and take as sample space the set of vectors $\Sigma = {1, 2,. . ., q}^V$, $V$ is the set of nodes of the graph $G$. So, each vertex of $G$ may be in any of $q$ states. For an edge $e = \langle x, y \rangle$ and a configuration $\sigma = (\sigma_x : x \in V) \in \Sigma$, we write $\delta_e(\sigma) = \delta_{\sigma x ,\sigma y}$, where $\delta_{i,j}$ is the Kronecker delta. 
The relevant probability measure is given by: $$\pi_{\beta,q} (\sigma) = \frac{1}{Z_p} exp(-\beta H'(\sigma)),$$ for $\sigma \in \Sigma$, where $Z_P = Z_P(\beta, q)$ is the appropriate normalizing constant and the Hamiltonian $H'$ is given by: $$ H'(\sigma) = - \Sigma_{e=\langle x, y \rangle} \delta_e (\sigma) $$

\subsection{Adaptation as a graph clustering quality measuring function}
Communities or clusters \cite{community_detection} generally accepted as groups of close distance or densely interconnected nodes and far distance or sparsely connected with other communities. Hence, as a quality measuring function, it should satisfy
\begin{itemize}
    \item reward internal links within the same community and non-links between different communities
(in the same spin state)
    \item vice versa, penalize non-links between nodes in the same community and links between communities 

\end{itemize}
Connecting the analogy of spin state and cluster, with $\delta_{ij}$ being the Kronecker delta, the Potts Model quality function is derived naturally as (negative sign was added for conventional minimizing loss function)
\begin{align*}
    \mathcal{H(\{\sigma\})} &= -\sum_{i\neq j}a_{ij}\underbrace{A_{ij}\delta(\sigma_i,\sigma_j)} _\text{internal links} + \sum_{i\neq j}b_{ij}\underbrace{(1-A_{ij})\delta(\sigma_i,\sigma_j)}_\text{internal non-links} \\
    &+ \sum_{i\neq j}c_{ij}\underbrace{A_{ij}(1-\delta(\sigma_i,\sigma_j))} _\text{external links} - \sum_{i\neq j}d_{ij}\underbrace{(1-A_{ij})(1-\delta(\sigma_i,\sigma_j))}_\text{external non-links} 
\end{align*}
    
with $\sigma_i \in \{1,2,...,q\}$  denotes the spin state (or group index) of node $i$ in the graph $G(V,E)$. \\
With the general assumption that links and non-links are each weighted equally, regardless whether they are external or internal $a_{ij} = c_{ij}$ and $b_{ij} = d_{ij}$, we can rewrite $\mathcal{H(\{\sigma\})}$ as the Hamiltonian
\begin{equation}
    \mathcal{H(\{\sigma\})} = -\sum_{i\neq j}\{ a_{ij} A_{ij} - b_{ij}(1-A_{ij})\} \delta(\sigma_i,\sigma_j)
    \label{hal}
\end{equation}
A choice of
weights $a_{ij}, b_{ij}$ will make the adjustment of contribution
of links and non-links easier by a change of parameter and help formulate the null model such that the partition is compared with. A common choice is a random null configuration model \cite{random_cluster} with a non-negative resolution parameter $\gamma$ by \cite{RB} $a_{ij} = 1 - \gamma p_{ij}$ 
and $bij = \gamma p_{ij}$ , where $p_{ij}$ denotes the probability that a link exists between node i and j, normalized, such
that $\sum_{i\neq j} p_{ij} = 2m$. In the case of $\gamma = 1$, which means the total amount of energy that can 
possibly be contributed by links and non-links is equal then the equation  \ref{hal} can be reduced to 
\begin{align*}
    \mathcal{H(\{\sigma\})} &= -\sum_{i\neq j}\{A_{ij} - \gamma p_{ij}\} \delta(\sigma_i,\sigma_j) \\
    &= -\sum_{c}\sum_{i\neq j}\{A_{ij} - \gamma p_{ij}\} \delta(\sigma_i,c)\delta(\sigma_j,c) 
\end{align*}
Rewrite the Hamiltonian given the expected number of edges $\{e_c\}_{p_{ij}} = \sum_{i\neq j} p_{ij}$ and probability $p_{ij} = \frac{k_i k_j}{2m}$ of the null configuration model \cite{null_model} \cite{perry2012null}
\begin{align}
    \mathcal{H(\{\sigma\})} &=  \frac{1}{2m}\sum_{c}[e_c - \gamma\frac{k_c^2}{2m} ] \label{modularity} \\
     &= \frac{1}{2m}\sum_{i \neq j}\{A_{ij} - \gamma \frac{k_ik_j}{2m}\} \delta(\sigma_i,\sigma_j)
\end{align}
Minimizing Hamiltonian corresponds to a partition of desirable characteristics. However, a minimum is not necessarily unique and better in general sense to a non minimum. Additionally, the choice of $\gamma$ has definite impact on community structure in which has to be chosen carefully. The degree $d_i$ of node $i$ is the number of connections from $V$ to $i$, the vector $\mathbf{d} = [d]_i$ contains the degrees of all the nodes in the graph. 
\subsection{Spectral Form Optimization}
In the discussion of complexity in the task of optimizing the spectral form of modularity in \cite{spectral_modularity_optimization} and \cite{dmon}, the problem is proven to be \textbf{NP-hard} and a relaxation version is empirically shown to be solved efficiently with a soft cluster matrix $\mathbf{C} \in \mathbb{R}^{n\times k}$ be the cluster assignment matrix and $d$ be the degree vector. Then, with matrix $\mathbf{P}$ defined as
\begin{align}
    \mathbf{P} &=\mathbf{A} - \gamma\frac{\mathbf{d}\mathbf{d}^T}{2m} \\
    \mathbf{P}x &= \mathbf{A}\mathbf{x} - \gamma\frac{\mathbf{d}\mathbf{x}\mathbf{d}^T}{2m} \label{mod_matrix}
\end{align}
Then Hamiltonian can be reformulated in matrix form 
\begin{equation}
    \mathcal{H} = \frac{1}{2m}Tr(\mathbf{C}^T\mathbf{P}\mathbf{C}) \label{loss}
\end{equation}
\section{GNNs for graph clustering}
\subsection{Previous work}
Recently, several architecture settings have been proposed in the literature
\begin{itemize}
    \item  Adaptive Graph Convolution (AGC),  Deep Attentional Embedded Graph Clustering (DAEGC) \cite{agc, daegc}: learn the embedding of node features based on convolutional and attentional GNN structure respectively.
    \item Deep Graph Infomax (DGI) \cite{dgi}: learn the embedding by maximizing mutual information
    \item Neural Overlapping Community Detection
(NOCD) \cite{nocd}: combine the 
power of GNNs and the Bernoulli–Poisson probabilistic model under reconstruction loss.
    \item Differential Pooling (DiffPool) \cite{diffpool} : is one of the first attempts developing a pooling layer that use message passing to learn and end-end unsupervised cluster matrix that relies on minimizing the entropy of the assignment and link prediction loss.
    \item MinCutPool\cite{mincut}: utilizes K-way normalized min cut problem as a optimize measuring function to find the interest partitions
    \item DMoN \cite{dmon}: learns the cluster assignment by optimizing modularity function
    \item Top-k \cite{topk}: learns an embedding vector to obtain a score from each node. The nodes with k highest scores are kept and the rest are dropped from the graph.
\end{itemize}
We can divide the recent GNN clustering algorithms into 2 classes. The first are algorithms that generate embeddings based on node feature and adjacency matrix. Afterwards, the embeddings are clustered
with k-means algorithm. The candidates for this class are AGC, DAEGC, DGI. The second learns the assignment matrix from end-end either by gate keeping strategy (Top-k) or optimizing for a global function (DMoN, MinCutPool, DiffPool). \\
Additionally, there have been efforts developing unsupervised algorithm for community detection that optimize Potts model loss function or modularity commonly, in a greedy fashion
\begin{itemize}
    \item Louvain \cite{louvain}: optimizing modularity in theoretical sense will results in the best possible parition. But going through all the combination to find the best modularity is expensive and impractical. Louvain is a heuristic approach to solve that problem.
    \item Leiden \cite{leiden}: an recent attempt to solve the issue of finding disconnected community. Leiden introduces one more phase into the system, refinement of partition.  Communities detected from the first modularity optimizing phase may split into smaller partitions in the second phase, inheritly solving the problem of finding small communities (resolution limit)
    \item Constant Potts Model \cite{cpm}: the layout of the algorithm is same as Louvain and optimizing constant Potts Model function instead of modularity.
All the above methods are extremely efficient and well studied. However, they are limited only with the graph structured and do not take the node features into the account for partitioning. 
\end{itemize}

\subsection{Graph Neural Networks}
The recent advance of adapting neural networks onto graph structured data is built on the message passing paradigm that extract the feature information of the local neighborhood. The feature aggregated is passed through a nonlinear transformation. In particular, the message passing architecture is described as 
\begin{equation}
    \mathbf{X}^{t+1} = \mathbf{MP}(\mathbf{A},\mathbf{X})
\end{equation}
 with $\mathbf{X}^{t+1}, \mathbf{X}^{t}$ being the output and input node features of $t$ message passing layer respectively. \\
There are several modification in the realm on GNN. A popular variant is   Graph Convolutional Networks (GCN) \cite{gcn}  uses implements the message passing function with a combination of linear transformations and ReLU function for normalized adjacency matrix $\mathbf{\bar{A}} = \mathbf{D}^{-1/2}\mathbf{A}\mathbf{D}^{-1/2}$
\begin{equation}
    \mathbf{X}^{t+1} = ReLU (\mathbf{\bar{A}}\mathbf{X}^{t}\mathbf{W})
\end{equation}
In this work, we employ the GCN layer \cite{gcn} as the setting for feature embedding encoder with skip layer going through  selu \cite{selu} activation function for better convergent and identitical to DMoN settings 
\begin{equation}
    \mathbf{X}^{t+1} = SeLU (\mathbf{\bar{A}}\mathbf{X}^{t}\mathbf{W} + \mathbf{W}_{skip})
\end{equation}
\section{Potts Model Networks (PMN)}
In this section, we discuss the Potts Model Networks (PMN) inspired by the modularity optimization from \cite{dmon} and the analysis and effort to overcome resolution limit \cite{cpm}. Moreover, the introduction of resolution parameter in the pooling layer and loss function provided a competitive advantage of flexibility in adapting the Potts Model loss function \ref{loss} onto the graph structure inherited by the training data.

Potts Model Networks (PMN) is a GNN layer that takes the normalized adjacency matrix $\mathbf{\bar{A}} = \mathbf{D}^{-1/2}\mathbf{A}\mathbf{D}^{-1/2}$ and obtains the soft cluster matrix $\mathbf{C}$ with $k$ being the max number of clusters
\begin{equation}
    \mathbf{C} = softmax(MP(\mathbf{\bar{A}}, \mathbf{X}), \mathbf{C} \in \mathbb{R}^{n\times k}
\end{equation}
The choice of message passing paradigm can be any kind of differentiable suitable function, the use of GCN (possibly multi-layers) in this work is a set up to compare directly with DMoN to analyze the difference of optimizing Potts Model function versus modularity. 
\subsection{Loss function}
The proposed loss function is an aggregation of the Hamiltonian \ref{hal} $\mathcal{L}_\mathcal{H}$, the collapse regularization \cite{dmon} $\mathcal{L}_c$, and the resolution parameter normalization $\mathcal{L}_\gamma$
\begin{align*}
    \mathcal{L}_{PMN}&= \mathcal{L}_\mathcal{H} + \mathcal{L}_c + \mathcal{L}_\gamma \\
    \mathcal{L}_\mathcal{H} &= \frac{1}{2m}Tr(\mathbf{C}^T\mathbf{P}\mathbf{C}) \\
    \mathcal{L}_c &= \frac{k}{\sqrt{n}}\normx{\sum_i\mathbf{C}_i^T}_F -1 \\
    \mathcal{L}_\gamma &= \normx{\gamma - \gamma_{max}}
\end{align*}
The introduction of $\gamma$ as a trained parameter is to motivate the PMN to learn the best resolution given the null configuration model. As discussed in \cite{leiden}, the influence $\gamma$ is interpreted as a threshold for detected communities. The inner-cluster density is filtered to be at least $\gamma$ while intra-cluster density should be lower than $\gamma$. The higher the resolution parameter is, the larger number of communities would be detected \cite{RB}. The resolution parameter $\gamma$ is first introduced in \cite{laplacian} to address a major problem with using Modularity as a global optimization function called resolution limit \cite{resolutionlimit}.
\subsection{Resolution-limit}
\cite{resolutionlimit} shows that modularity as quality measuring function inheritly has a filter scale that depends on the size of the network. Communities that are smaller than this filter scale may not be detected even if they are complete and fully connected. The reason lies on the use of modularity as a sum of modules to be a global metric for optimization. Finding the best modularity is a trade-off between the number of communities and the modular value of each term. An increase  in number of communities doesn't necessarily yield an increase in the global modularity because modular value for each community will be smaller. 
\subsection{Optimizing resolution}
\begin{figure}
	\centering
	\includegraphics[width=0.5\linewidth]{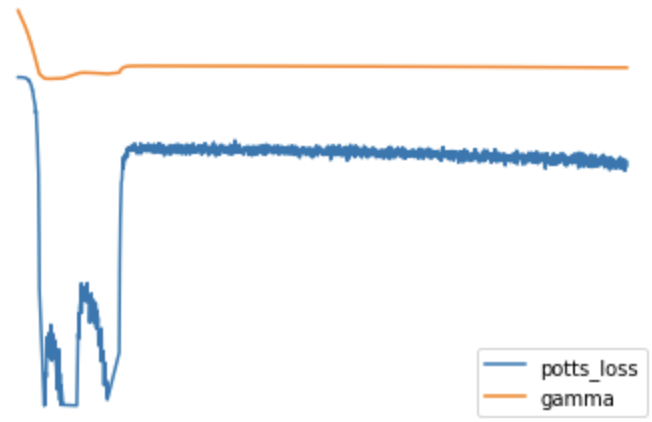}
	\caption{The convergence of potts loss and gamma in the process of optimization, the gamma converges into a stable value leads to the stable convergence of Potts loss}
	\label{fig:gamma_convergence}
\end{figure}

A limit of the framework that coarsen node features and edge link into a cluster assignment matrix is the number of maximum clusters have to be defined before the training process. There is no indication information for the task of choosing the suitable $k$ number of clusters. Modularity, as described above, does suffer from resolution limit and not necessarily a good indication to tune the number of clusters. For Potts Model, the resolution parameter $\gamma$ plays the role as an indication of granularity of communities in a heuristic greedy approach. Applying Potts Model as a loss function and involving $\gamma$ as a training parameter provide two competitive edges:
\begin{itemize}
    \item Indication to tune parameter $k$ - number of clusters
    \item Adaptation of the loss function onto the graph sparsity and structure
\end{itemize}
As shown in \ref{fig:gamma_convergence}, the PMN $\gamma$ variable stabilized its training process first leading the convergence of the optimal Potts loss convergence. This demonstrates the ability to adapt the loss function onto the graph structure of each dataset.

\section{Experiments \& results}
\textbf{Benchmark datasets:} we borrow the benchmark results from \cite{dmon} on 10 real-world datasets from citation networks Cora, CiteSeer, and PubMed \cite{plantoid}, coauthors networks based on Microsoft Academic Graph (CS, Physics, Med, Chem, Engineering) \cite{nocd,coauthor,microsoft}, to Amazon co-purchase graph (Photo, Computers) \cite{amazon}. The features of nodes are bag-of-word for abstract, paper keywords, or reviews. The corresponding labels shows topic of papers, fields of study, or product category. 
\\
\textbf{Metrics:} we use the standard metrics to measure quality of cluster. On the graph side, we employ conductance (C)\cite{conductance} which measures the edge volume that points outside the cluster, and modularity (Q)\cite{newman2006modularity} against existing benchmark. For the correlation with ground truth label, we use normalized mutual information score (NMI) and F1 score \cite{nmi}. Conductance is ranked the lower the better, while the rest of the metrics, the higher the better. We will scale all metric to the range of $100$ for better aesthetic and comparison. 
\\

\begin{wraptable}{r}{7.5cm}
\centering
\begin{tabular}{|l|l|l|l|l|}
\hline
{\color[HTML]{242424} Dataset}       & |V|                          & {\color[HTML]{4F4F4F} |E|}    & {\color[HTML]{4F4F4F} |X|}  & |Y|                       \\ \hline
{\color[HTML]{242424} Cora}          & {\color[HTML]{242424} 2708}  & {\color[HTML]{242424} 5278}   & {\color[HTML]{242424} 1433} & {\color[HTML]{131313} 7}  \\ \hline
{\color[HTML]{242424} CiteSeer}      & {\color[HTML]{242424} 3327}  & {\color[HTML]{131313} 4614}   & {\color[HTML]{242424} 3703} & {\color[HTML]{242424} 6}  \\ \hline
{\color[HTML]{242424} Pubmed}        & {\color[HTML]{242424} 19717} & {\color[HTML]{242424} 44325}  & {\color[HTML]{242424} 500}  & {\color[HTML]{242424} 3}  \\ \hline
{\color[HTML]{242424} Amazon Computers}     & {\color[HTML]{242424} 13752} & {\color[HTML]{131313} 143604} & {\color[HTML]{131313} 767}  & {\color[HTML]{131313} 10} \\ \hline
{\color[HTML]{242424} Amazon Photo}  & {\color[HTML]{131313} 7650}  & {\color[HTML]{131313} 71831}  & {\color[HTML]{131313} 745}  & {\color[HTML]{242424} 8}  \\ \hline
{\color[HTML]{242424} Coauthor Eng}  & {\color[HTML]{242424} 14927} & {\color[HTML]{242424} 49305}  & {\color[HTML]{131313} 4839} & {\color[HTML]{131313} 16} \\ \hline
{\color[HTML]{242424} Coauthor CS}   & {\color[HTML]{242424} 18333} & {\color[HTML]{242424} 81894}  & {\color[HTML]{242424} 6805} & {\color[HTML]{131313} 15} \\ \hline
{\color[HTML]{242424} Coauthor Phys} & {\color[HTML]{242424} 34493} & {\color[HTML]{242424} 247962} & {\color[HTML]{242424} 8415} & {\color[HTML]{242424} 5}  \\ \hline
{\color[HTML]{242424} Coauthor Chem} & {\color[HTML]{242424} 35409} & {\color[HTML]{131313} 157358} & {\color[HTML]{131313} 4877} & {\color[HTML]{131313} 14} \\ \hline
{\color[HTML]{242424} Coauthor Med}  & {\color[HTML]{242424} 63282} & {\color[HTML]{242424} 810314} & {\color[HTML]{242424} 5538} & {\color[HTML]{131313} 17} \\ \hline
\end{tabular}
\caption{Detail information of the benchmark datasets: number of
nodes $|V|$, number of edges $|E|$, dimension of the node features $|X|$, and number of labels $|Y|$.}
\label{tab:dataset}
\end{wraptable}

\textbf{Model configurations:} in our experiments, the GNN model used for PMN is built on top of the DMoN architecture, and all parameters are kept the same for direct comparison.The baseline results that we compared with are from \cite{dmon} results. The encoder includes $1$ layer with $64$ units and the max number of clusters $k$ equals to $16$. Dropout is kept at $0.5$ and the maximum $\gamma$ is fixed at $5$. The initialization of weight matrices are $\mathcal{N}(0,1)$ and $\gamma_{init} = 1$. The collapse and $\gamma$ regularization are $1,0.01$ respectively. We run the experiments for 4 models MinCut, MinCut with Orthogonal loss \cite{mincut}, DMoN \cite{dmon}, and PMN. All the parameters of MinCut and DMoN are the same as described above, we also borrow the results from \cite{dmon} to demonstrate the comprehensive comparison between all methods. The results are averaged over 10 runs of random seeds. \\
\textbf{Results:} the performance of Potts and our runs of DMoN, Ortho, MinCut are shown in table \ref{amazon-table},\ref{coauthor-table},\ref{plantoid}. The results display strong and consistent performance of PMN over 10 datasets. PMN consistently perform worse in Modularity (Q) while better in other metrics comparing to DMoN with same configuration. Notably, PMN performs extremely well on coauthorship dataset \cite{coauthor}. PMN achieves state-of-the-art conductance, F1, NMI on Cora, PubMed, and all Coauthor datasets. The lack of performance in modularity measure expected because PMN is not optimizing for modularity directly, it finds the best resolution for the graph data as well as minimizing the Potts loss function \ref{loss}. Because of that, we achieve significantly better results in F1, NMI, conductance score in Coauthor Phys with a $200\%$ increasing from the second best of $42.9 (F1), 23.5 (C)$ to $88.2, 5.5$ respectively. In general, DMoN is better at maximizing modularity, $10-30\%$ better than PMN consistantly. In our runs, there is difficulty to reproduce all the results from DMoN \cite{dmon} when following the configuration of parameter settings. Even so, PMN still show better result most of the time except for the citation datasets. Overall, PMN performs dominantly better than its counter part DMoN except modularity measurement. 



\begin{table}[!h]
\centering
\resizebox{\columnwidth}{!}{%
\begin{tabular}{|l|llll|llll|llll|}
\hline
{\color[HTML]{242424} } &
  \multicolumn{4}{l|}{Coauthor Phys} &
  \multicolumn{4}{l|}{Coauthor Chem} &
  \multicolumn{4}{l|}{Coauthor Med} \\ \cline{2-13} 
{\color[HTML]{242424} } &
  \multicolumn{2}{l|}{Graph} &
  \multicolumn{2}{l|}{Labels} &
  \multicolumn{2}{l|}{Graph} &
  \multicolumn{2}{l|}{Labels} &
  \multicolumn{2}{l|}{Graph} &
  \multicolumn{2}{l|}{Labels} \\ \cline{2-13} 
\multirow{-3}{*}{{\color[HTML]{242424} Methods}} &
  \multicolumn{1}{l|}{C} &
  \multicolumn{1}{l|}{Q} &
  \multicolumn{1}{l|}{NMI} &
  F1 &
  \multicolumn{1}{l|}{C} &
  \multicolumn{1}{l|}{Q} &
  \multicolumn{1}{l|}{NMI} &
  F1 &
  \multicolumn{1}{l|}{C} &
  \multicolumn{1}{l|}{Q} &
  \multicolumn{1}{l|}{NMI} &
  F1 \\ \hline
k-m(feat) &
  \multicolumn{1}{l|}{57.0} &
  \multicolumn{1}{l|}{19.4} &
  \multicolumn{1}{l|}{30.6} &
  42.9 &
  \multicolumn{1}{l|}{42.9} &
  \multicolumn{1}{l|}{18.2} &
  \multicolumn{1}{l|}{13.9} &
  35.l &
  \multicolumn{1}{l|}{54.7} &
  \multicolumn{1}{l|}{19.3} &
  \multicolumn{1}{l|}{11.8} &
  31.7 \\ \hline
SBM &
  \multicolumn{1}{l|}{25.9} &
  \multicolumn{1}{l|}{\textbf{66.9}} &
  \multicolumn{1}{l|}{45.4} &
  30.4 &
  \multicolumn{1}{l|}{18.4} &
  \multicolumn{1}{l|}{74.6} &
  \multicolumn{1}{l|}{25.4} &
  25.0 &
  \multicolumn{1}{l|}{21.1} &
  \multicolumn{1}{l|}{72.0} &
  \multicolumn{1}{l|}{36.l} &
  31.1 \\ \hline
DeepWalk &
  \multicolumn{1}{l|}{44.7} &
  \multicolumn{1}{l|}{47.0} &
  \multicolumn{1}{l|}{43.5} &
  24.3 &
  \multicolumn{1}{l|}{14.0} &
  \multicolumn{1}{l|}{74.8} &
  \multicolumn{1}{l|}{36.5} &
  33.8 &
  \multicolumn{1}{l|}{16.6} &
  \multicolumn{1}{l|}{72.1} &
  \multicolumn{1}{l|}{43.l} &
  39.4 \\ \hline
SDCN &
  \multicolumn{1}{l|}{32.l} &
  \multicolumn{1}{l|}{52.8} &
  \multicolumn{1}{l|}{50.4} &
  39.9 &
  \multicolumn{1}{l|}{29.9} &
  \multicolumn{1}{l|}{58.7} &
  \multicolumn{1}{l|}{33.3} &
  32.8 &
  \multicolumn{1}{l|}{34.8} &
  \multicolumn{1}{l|}{54.2} &
  \multicolumn{1}{l|}{25.2} &
  26.5 \\ \hline
DGI &
  \multicolumn{1}{l|}{38.6} &
  \multicolumn{1}{l|}{51.2} &
  \multicolumn{1}{l|}{51.0} &
  30.6 &
  \multicolumn{1}{l|}{31.6} &
  \multicolumn{1}{l|}{60.6} &
  \multicolumn{1}{l|}{40.8} &
  32.9 &
  \multicolumn{1}{l|}{35.7} &
  \multicolumn{1}{l|}{56.5} &
  \multicolumn{1}{l|}{34.8} &
  27.7 \\ \hline
NOCD &
  \multicolumn{1}{l|}{25.7} &
  \multicolumn{1}{l|}{65.5} &
  \multicolumn{1}{l|}{51.9} &
  28.7 &
  \multicolumn{1}{l|}{19.2} &
  \multicolumn{1}{l|}{73.l} &
  \multicolumn{1}{l|}{43.1} &
  40.l &
  \multicolumn{1}{l|}{22.0} &
  \multicolumn{1}{l|}{69.7} &
  \multicolumn{1}{l|}{42.5} &
  37.6 \\ \hline
MinCut &
  \multicolumn{1}{l|}{29.7} &
  \multicolumn{1}{l|}{63.1} &
  \multicolumn{1}{l|}{42.3} &
  32.4 &
  \multicolumn{1}{l|}{21.4} &
  \multicolumn{1}{l|}{70.1} &
  \multicolumn{1}{l|}{41.5} &
  33.5 &
  \multicolumn{1}{l|}{24.2} &
  \multicolumn{1}{l|}{70.1} &
  \multicolumn{1}{l|}{43.1} &
  33.4 \\ \hline
Ortho &
  \multicolumn{1}{l|}{31.1} &
  \multicolumn{1}{l|}{63.5} &
  \multicolumn{1}{l|}{46.4} &
  35.8 &
  \multicolumn{1}{l|}{21.6} &
  \multicolumn{1}{l|}{69.2} &
  \multicolumn{1}{l|}{39.1} &
  29.3 &
  \multicolumn{1}{l|}{21.5} &
  \multicolumn{1}{l|}{68.3} &
  \multicolumn{1}{l|}{39.5} &
  31.5 \\ \hline
DMoN &
  \multicolumn{1}{l|}{23.5} &
  \multicolumn{1}{l|}{66.3} &
  \multicolumn{1}{l|}{53.8} &
  37.8 &
  \multicolumn{1}{l|}{21.8} &
  \multicolumn{1}{l|}{71.2} &
  \multicolumn{1}{l|}{46.3} &
  44.9 &
  \multicolumn{1}{l|}{22.2} &
  \multicolumn{1}{l|}{72.2} &
  \multicolumn{1}{l|}{51.5} &
  50.9 \\ \hline
Potts &
  \multicolumn{1}{l|}{\textbf{5.5}} &
  \multicolumn{1}{l|}{49.5} &
  \multicolumn{1}{l|}{\textbf{72.2}} &
  \textbf{88.2} &
  \multicolumn{1}{l|}{\textbf{11.3}} &
  \multicolumn{1}{l|}{\textbf{75.7}} &
  \multicolumn{1}{l|}{\textbf{58.1}} &
  \textbf{57.1} &
  \multicolumn{1}{l|}{\textbf{14.9}} &
  \multicolumn{1}{l|}{\textbf{72.6}} &
  \multicolumn{1}{l|}{\textbf{60.4}} &
  \textbf{59.7} \\ \hline
\end{tabular}%
}
\caption{Performance of benchmark models on medium size graph datasets}
\label{coauthor-table}
\end{table}
\begin{table}[!h]
\centering
\resizebox{\columnwidth}{!}{%
\begin{tabular}{|l|llll|llll|llll|}
\hline
{\color[HTML]{242424} } &
  \multicolumn{4}{c|}{Cora} &
  \multicolumn{4}{c|}{Citeseer} &
  \multicolumn{4}{c|}{Pubmed} \\ \cline{2-13} 
\multirow{-2}{*}{{\color[HTML]{242424} Methods}} &
  \multicolumn{2}{c|}{graph} &
  \multicolumn{2}{c|}{labels} &
  \multicolumn{2}{c|}{graph} &
  \multicolumn{2}{c|}{labels} &
  \multicolumn{2}{c|}{graph} &
  \multicolumn{2}{c|}{labels} \\ \hline
Metrics &
  \multicolumn{1}{c|}{C} &
  \multicolumn{1}{c|}{Q} &
  \multicolumn{1}{c|}{NMI} &
  \multicolumn{1}{c|}{F1} &
  \multicolumn{1}{c|}{C} &
  \multicolumn{1}{c|}{Q} &
  \multicolumn{1}{c|}{NMI} &
  \multicolumn{1}{c|}{F1} &
  \multicolumn{1}{c|}{C} &
  \multicolumn{1}{c|}{Q} &
  \multicolumn{1}{c|}{NMI} &
  \multicolumn{1}{c|}{F1} \\ \hline
k-m(feat) &
  \multicolumn{1}{l|}{61.7} &
  \multicolumn{1}{l|}{19.8} &
  \multicolumn{1}{l|}{18.5} &
  27.0 &
  \multicolumn{1}{l|}{60.5} &
  \multicolumn{1}{l|}{30.3} &
  \multicolumn{1}{l|}{24.5} &
  29.2 &
  \multicolumn{1}{l|}{55.8} &
  \multicolumn{1}{l|}{33.4} &
  \multicolumn{1}{l|}{19.4} &
  24.4 \\ \hline
SBM &
  \multicolumn{1}{l|}{15.4} &
  \multicolumn{1}{l|}{\textbf{77.3}} &
  \multicolumn{1}{l|}{36.2} &
  30.2 &
  \multicolumn{1}{l|}{14.2} &
  \multicolumn{1}{l|}{78.1} &
  \multicolumn{1}{l|}{15.3} &
  19.1 &
  \multicolumn{1}{l|}{39.0} &
  \multicolumn{1}{l|}{53.5} &
  \multicolumn{1}{l|}{16.4} &
  16.7 \\ \hline
DeepWalk &
  \multicolumn{1}{l|}{62.1} &
  \multicolumn{1}{l|}{30.7} &
  \multicolumn{1}{l|}{24.3} &
  24.8 &
  \multicolumn{1}{l|}{68.1} &
  \multicolumn{1}{l|}{24.3} &
  \multicolumn{1}{l|}{27.6} &
  24.8 &
  \multicolumn{1}{l|}{16.6} &
  \multicolumn{1}{l|}{\textbf{75.3}} &
  \multicolumn{1}{l|}{22.9} &
  17.2 \\ \hline
AGC &
  \multicolumn{1}{l|}{48.9} &
  \multicolumn{1}{l|}{43.2} &
  \multicolumn{1}{l|}{34.1} &
  28.9 &
  \multicolumn{1}{l|}{41.9} &
  \multicolumn{1}{l|}{50.0} &
  \multicolumn{1}{l|}{25.5} &
  27.5 &
  \multicolumn{1}{l|}{44.9} &
  \multicolumn{1}{l|}{46.8} &
  \multicolumn{1}{l|}{18.2} &
  18.4 \\ \hline
SDCN &
  \multicolumn{1}{l|}{37.5} &
  \multicolumn{1}{l|}{50.8} &
  \multicolumn{1}{l|}{27.9} &
  29.9 &
  \multicolumn{1}{l|}{20.0} &
  \multicolumn{1}{l|}{62.3} &
  \multicolumn{1}{l|}{\textbf{31.4}} &
  \textbf{41.9} &
  \multicolumn{1}{l|}{22.4} &
  \multicolumn{1}{l|}{50.3} &
  \multicolumn{1}{l|}{19.5} &
  29.9 \\ \hline
DAEGC &
  \multicolumn{1}{l|}{56.8} &
  \multicolumn{1}{l|}{33.5} &
  \multicolumn{1}{l|}{8.3} &
  13.6 &
  \multicolumn{1}{l|}{47.6} &
  \multicolumn{1}{l|}{36.4} &
  \multicolumn{1}{l|}{4.3} &
  18.0 &
  \multicolumn{1}{l|}{53.6} &
  \multicolumn{1}{l|}{37.5} &
  \multicolumn{1}{l|}{4.4} &
  11.6 \\ \hline
DGI &
  \multicolumn{1}{l|}{28.0} &
  \multicolumn{1}{l|}{64.0} &
  \multicolumn{1}{l|}{52.7} &
  40.1 &
  \multicolumn{1}{l|}{17.5} &
  \multicolumn{1}{l|}{73.7} &
  \multicolumn{1}{l|}{40.4} &
  39.4 &
  \multicolumn{1}{l|}{82.9} &
  \multicolumn{1}{l|}{9.6} &
  \multicolumn{1}{l|}{22.0} &
  26.4 \\ \hline
NOCD &
  \multicolumn{1}{l|}{14.7} &
  \multicolumn{1}{l|}{78.3} &
  \multicolumn{1}{l|}{46.3} &
  36.7 &
  \multicolumn{1}{l|}{6.8} &
  \multicolumn{1}{l|}{84.4} &
  \multicolumn{1}{l|}{20.0} &
  24.1 &
  \multicolumn{1}{l|}{21.7} &
  \multicolumn{1}{l|}{69.6} &
  \multicolumn{1}{l|}{25.5} &
  20.8 \\ \hline
DiffPool &
  \multicolumn{1}{l|}{26.1} &
  \multicolumn{1}{l|}{66.3} &
  \multicolumn{1}{l|}{32.9} &
  34.4 &
  \multicolumn{1}{l|}{26.0} &
  \multicolumn{1}{l|}{63.4} &
  \multicolumn{1}{l|}{20.0} &
  23.5 &
  \multicolumn{1}{l|}{32.9} &
  \multicolumn{1}{l|}{56.8} &
  \multicolumn{1}{l|}{20.2} &
  26.3 \\ \hline
MinCut &
  \multicolumn{1}{l|}{29.3} &
  \multicolumn{1}{l|}{71.5} &
  \multicolumn{1}{l|}{30.1} &
  25.0 &
  \multicolumn{1}{l|}{14.1} &
  \multicolumn{1}{l|}{82.2} &
  \multicolumn{1}{l|}{25.9} &
  20.1 &
  \multicolumn{1}{l|}{22.3} &
  \multicolumn{1}{l|}{64.5} &
  \multicolumn{1}{l|}{24.1} &
  28.5 \\ \hline
Ortho &
  \multicolumn{1}{l|}{19.2} &
  \multicolumn{1}{l|}{65.6} &
  \multicolumn{1}{l|}{29.4} &
  26.6 &
  \multicolumn{1}{l|}{15.4} &
  \multicolumn{1}{l|}{79.2} &
  \multicolumn{1}{l|}{30.1} &
  19.2 &
  \multicolumn{1}{l|}{47.7} &
  \multicolumn{1}{l|}{38.2} &
  \multicolumn{1}{l|}{21.0} &
  18.4 \\ \hline
DMoN &
  \multicolumn{1}{l|}{11.8} &
  \multicolumn{1}{l|}{75.8} &
  \multicolumn{1}{l|}{45.6} &
  35.9 &
  \multicolumn{1}{l|}{6.9} &
  \multicolumn{1}{l|}{\textbf{83.2}} &
  \multicolumn{1}{l|}{27.9} &
  31.4 &
  \multicolumn{1}{l|}{17.2} &
  \multicolumn{1}{l|}{68.3} &
  \multicolumn{1}{l|}{30.2} &
  40.1 \\ \hline
Potts &
  \multicolumn{1}{l|}{\textbf{5.5}} &
  \multicolumn{1}{l|}{57.8} &
  \multicolumn{1}{l|}{\textbf{49.7}} &
  \textbf{54.7} &
  \multicolumn{1}{l|}{6.0} &
  \multicolumn{1}{l|}{81.2} &
  \multicolumn{1}{l|}{29.2} &
  36.9 &
  \multicolumn{1}{l|}{\textbf{7.4}} &
  \multicolumn{1}{l|}{59.4} &
  \multicolumn{1}{l|}{\textbf{32.4}} &
  \textbf{56.3} \\ \hline
\end{tabular}%
}
\caption{Citation network dataset performance}
\label{plantoid}
\end{table}

\begin{table}[!h]
\centering
\caption{Large graph dataset performance}
\resizebox{\columnwidth}{!}{%
\begin{tabular}{|l|llllllll|llll|llll|}
\hline
{\color[HTML]{242424} } &
  \multicolumn{4}{l|}{Amazon PC} &
  \multicolumn{4}{l|}{Amazon Photo} &
  \multicolumn{4}{l|}{Coauthor CS} &
  \multicolumn{4}{l|}{Coauthor Eng} \\ \cline{2-17} 
\multirow{-2}{*}{{\color[HTML]{242424} Methods}} &
  \multicolumn{2}{l|}{Graph } &
  \multicolumn{2}{l|}{Label} &
  \multicolumn{2}{l|}{Graph} &
  \multicolumn{2}{l|}{Label} &
  \multicolumn{2}{l|}{Graph} &
  \multicolumn{2}{l|}{Label} &
  \multicolumn{2}{l|}{Graph} &
  \multicolumn{2}{l|}{Label} \\ \hline
 &
  \multicolumn{1}{l|}{C} &
  \multicolumn{1}{l|}{Q} &
  \multicolumn{1}{l|}{NMI} &
  \multicolumn{1}{l|}{F1} &
  \multicolumn{1}{l|}{C} &
  \multicolumn{1}{l|}{Q} &
  \multicolumn{1}{l|}{NMI} &
  F1 &
  \multicolumn{1}{l|}{C} &
  \multicolumn{1}{l|}{Q} &
  \multicolumn{1}{l|}{NMI} &
  F1 &
  \multicolumn{1}{l|}{C} &
  \multicolumn{1}{l|}{Q} &
  \multicolumn{1}{l|}{NMI} &
  F1 \\ \hline
k-m (feat) &
  \multicolumn{1}{l|}{84.5} &
  \multicolumn{1}{l|}{5.4} &
  \multicolumn{1}{l|}{21.1} &
  \multicolumn{1}{l|}{19.2} &
  \multicolumn{1}{l|}{79.6} &
  \multicolumn{1}{l|}{10.5} &
  \multicolumn{1}{l|}{28.8} &
  19.5 &
  \multicolumn{1}{l|}{49.1} &
  \multicolumn{1}{l|}{23.1} &
  \multicolumn{1}{l|}{35.7} &
  39.4 &
  \multicolumn{1}{l|}{42.7} &
  \multicolumn{1}{l|}{27.1} &
  \multicolumn{1}{l|}{24.5} &
  32.5 \\ \hline
SBM &
  \multicolumn{1}{l|}{31.0} &
  \multicolumn{1}{l|}{\textbf{60.8}} &
  \multicolumn{1}{l|}{48.4} &
  \multicolumn{1}{l|}{34.6} &
  \multicolumn{1}{l|}{18.6} &
  \multicolumn{1}{l|}{\textbf{72.7}} &
  \multicolumn{1}{l|}{59.3} &
  47.4 &
  \multicolumn{1}{l|}{20.3} &
  \multicolumn{1}{l|}{72.7} &
  \multicolumn{1}{l|}{58.0} &
  47.7 &
  \multicolumn{1}{l|}{15.8} &
  \multicolumn{1}{l|}{77.0} &
  \multicolumn{1}{l|}{33.3} &
  27.5 \\ \hline
DeepWalk &
  \multicolumn{1}{l|}{67.6} &
  \multicolumn{1}{l|}{11.8} &
  \multicolumn{1}{l|}{38.2} &
  \multicolumn{1}{l|}{22.7} &
  \multicolumn{1}{l|}{60.6} &
  \multicolumn{1}{l|}{22.9} &
  \multicolumn{1}{l|}{49.4} &
  33.8 &
  \multicolumn{1}{l|}{33.1} &
  \multicolumn{1}{l|}{59.4} &
  \multicolumn{1}{l|}{72.7} &
  61.2 &
  \multicolumn{1}{l|}{5.7} &
  \multicolumn{1}{l|}{67.4} &
  \multicolumn{1}{l|}{47.7} &
  50.0 \\ \hline
AGC &
  \multicolumn{1}{l|}{43.2} &
  \multicolumn{1}{l|}{42.8} &
  \multicolumn{1}{l|}{\textbf{51.3}} &
  \multicolumn{1}{l|}{35.3} &
  \multicolumn{1}{l|}{33.8} &
  \multicolumn{1}{l|}{55.9} &
  \multicolumn{1}{l|}{59.0} &
  44.2 &
  \multicolumn{1}{l|}{41.5} &
  \multicolumn{1}{l|}{40.1} &
  \multicolumn{1}{l|}{43.3} &
  31.9 &
  \multicolumn{1}{l|}{32.3} &
  \multicolumn{1}{l|}{46.4} &
  \multicolumn{1}{l|}{30.8} &
  31.2 \\ \hline
DAEGC &
  \multicolumn{1}{l|}{39.0} &
  \multicolumn{1}{l|}{43.3} &
  \multicolumn{1}{l|}{42.5} &
  \multicolumn{1}{l|}{37.3} &
  \multicolumn{1}{l|}{19.3} &
  \multicolumn{1}{l|}{58.0} &
  \multicolumn{1}{l|}{47.6} &
  45.0 &
  \multicolumn{1}{l|}{39.4} &
  \multicolumn{1}{l|}{49.1} &
  \multicolumn{1}{l|}{36.3} &
  32.4 &
  \multicolumn{1}{l|}{31.9} &
  \multicolumn{1}{l|}{50.9} &
  \multicolumn{1}{l|}{12.5} &
  26.1 \\ \hline
SDCN &
  \multicolumn{1}{l|}{25.1} &
  \multicolumn{1}{l|}{45.6} &
  \multicolumn{1}{l|}{24.9} &
  \multicolumn{1}{l|}{45.2} &
  \multicolumn{1}{l|}{19.7} &
  \multicolumn{1}{l|}{53.3} &
  \multicolumn{1}{l|}{41.7} &
  45.1 &
  \multicolumn{1}{l|}{33.0} &
  \multicolumn{1}{l|}{55.7} &
  \multicolumn{1}{l|}{59.3} &
  54.7 &
  \multicolumn{1}{l|}{21.8} &
  \multicolumn{1}{l|}{64.6} &
  \multicolumn{1}{l|}{45.3} &
  45.9 \\ \hline
DGI &
  \multicolumn{1}{l|}{61.9} &
  \multicolumn{1}{l|}{22.8} &
  \multicolumn{1}{l|}{22.6} &
  \multicolumn{1}{l|}{15.0} &
  \multicolumn{1}{l|}{51.5} &
  \multicolumn{1}{l|}{35.1} &
  \multicolumn{1}{l|}{33.4} &
  23.6 &
  \multicolumn{1}{l|}{35.1} &
  \multicolumn{1}{l|}{57.8} &
  \multicolumn{1}{l|}{64.6} &
  51.9 &
  \multicolumn{1}{l|}{29.3} &
  \multicolumn{1}{l|}{60.4} &
  \multicolumn{1}{l|}{49.7} &
  37.2 \\ \hline
NOCD &
  \multicolumn{1}{l|}{26.4} &
  \multicolumn{1}{l|}{59.0} &
  \multicolumn{1}{l|}{44.8} &
  \multicolumn{1}{l|}{37.8} &
  \multicolumn{1}{l|}{13.7} &
  \multicolumn{1}{l|}{70.1} &
  \multicolumn{1}{l|}{62.3} &
  60.2 &
  \multicolumn{1}{l|}{20.9} &
  \multicolumn{1}{l|}{72.2} &
  \multicolumn{1}{l|}{70.5} &
  56.4 &
  \multicolumn{1}{l|}{16.0} &
  \multicolumn{1}{l|}{75.6} &
  \multicolumn{1}{l|}{50.7} &
  35.4 \\ \hline
DiffPool &
  \multicolumn{1}{l|}{35.6} &
  \multicolumn{1}{l|}{30.4} &
  \multicolumn{1}{l|}{22.1} &
  \multicolumn{1}{l|}{38.3} &
  \multicolumn{1}{l|}{26.5} &
  \multicolumn{1}{l|}{46.8} &
  \multicolumn{1}{l|}{35.9} &
  41.8 &
  \multicolumn{1}{l|}{33.6} &
  \multicolumn{1}{l|}{59.3} &
  \multicolumn{1}{l|}{41.6} &
  34.4 &
  \multicolumn{1}{l|}{34.9} &
  \multicolumn{1}{l|}{55.0} &
  \multicolumn{1}{l|}{22.0} &
  21.8 \\ \hline
MinCut &
  \multicolumn{8}{c|}{did not converge} &
  \multicolumn{1}{l|}{20.1} &
  \multicolumn{1}{l|}{70.5} &
  \multicolumn{1}{l|}{63.2} &
  49.5 &
  \multicolumn{1}{l|}{25.3} &
  \multicolumn{1}{l|}{73.2} &
  \multicolumn{1}{l|}{45.2} &
  34.2 \\ \hline
Ortho &
  \multicolumn{8}{c|}{did not converge} &
  \multicolumn{1}{l|}{26.7} &
  \multicolumn{1}{l|}{65.7} &
  \multicolumn{1}{l|}{58.2} &
  45.3 &
  \multicolumn{1}{l|}{24.5} &
  \multicolumn{1}{l|}{75.4} &
  \multicolumn{1}{l|}{46.1} &
  32.2 \\ \hline
DMoN &
  \multicolumn{1}{l|}{18.0} &
  \multicolumn{1}{l|}{59.0} &
  \multicolumn{1}{l|}{49.3} &
  \multicolumn{1}{l|}{45.4} &
  \multicolumn{1}{l|}{21.1} &
  \multicolumn{1}{l|}{68.7} &
  \multicolumn{1}{l|}{58.9} &
  47.8 &
  \multicolumn{1}{l|}{17.9} &
  \multicolumn{1}{l|}{\textbf{73.3}} &
  \multicolumn{1}{l|}{69.5} &
  59.2 &
  \multicolumn{1}{l|}{14.4} &
  \multicolumn{1}{l|}{\textbf{77.2}} &
  \multicolumn{1}{l|}{57.8} &
  48.3 \\ \hline
Potts &
  \multicolumn{1}{l|}{\textbf{15.4}} &
  \multicolumn{1}{l|}{53.8} &
  \multicolumn{1}{l|}{44.4} &
  \multicolumn{1}{l|}{\textbf{49.8}} &
  \multicolumn{1}{l|}{\textbf{6.5}} &
  \multicolumn{1}{l|}{59.0} &
  \multicolumn{1}{l|}{\textbf{62.6}} &
  \textbf{59.2} &
  \multicolumn{1}{l|}{\textbf{12.6}} &
  \multicolumn{1}{l|}{72.9} &
  \multicolumn{1}{l|}{\textbf{75.2}} &
  \textbf{78.7} &
  \multicolumn{1}{l|}{\textbf{5.2}} &
  \multicolumn{1}{l|}{72.2} &
  \multicolumn{1}{l|}{\textbf{63.1}} &
  \textbf{69.5} \\ \hline
\end{tabular}%
}
\label{amazon-table}
\end{table}

\section{Conclusions}
In this paper, we introduced a generalized framework to address important limitations of modularity optimization and proposed a new trainable measurement of clustering quality. The Potts Model Network approach relies on the existing Potts model application in statistical mechanics for community detection. Moreover, we explore the performance of PMN model on 10 real life datasets and achieve desirable results comparing to existing pooling framework. \\
Future research direction could be developing an approach to utilize the trained $gamma$ to tune the number of clusters. Investigating on the effect of sparsity on the performance of Potts Model is promising based on the empirical results from our experiments. 
\section{Acknowledgement}
We would like to thank the SailPoint Technologies for the computational support.

\appendix
\section{Appendix}

\printbibliography

@inproceedings{gcn,
author = {Kipf, Thomas N. and Welling, Max},
title = {{Semi-Supervised Classification with Graph Convolutional Networks}},
booktitle = {ICLR},
year = {2017}
}

@article{dmon,
  title={Graph clustering with graph neural networks},
  author={Tsitsulin, Anton and Palowitch, John and Perozzi, Bryan and M{\"u}ller, Emmanuel},
  journal={arXiv preprint arXiv:2006.16904},
  year={2020}
}

@inproceedings{mincut,
  title={Spectral clustering with graph neural networks for graph pooling},
  author={Bianchi, Filippo Maria and Grattarola, Daniele and Alippi, Cesare},
  booktitle={International Conference on Machine Learning},
  pages={874--883},
  year={2020},
  organization={PMLR}
}

@article{spectral_modularity_optimization,
  title={Finding community structure in networks using the eigenvectors of matrices},
  author={Newman, Mark EJ},
  journal={Physical review E},
  volume={74},
  number={3},
  pages={036104},
  year={2006},
  publisher={APS}
}

@article{RB,
  title={Statistical mechanics of community detection},
  author={Reichardt, J{\"o}rg and Bornholdt, Stefan},
  journal={Physical review E},
  volume={74},
  number={1},
  pages={016110},
  year={2006},
  publisher={APS}
}

@article{null_model,
  title={The structure and function of complex networks},
  author={Newman, Mark EJ},
  journal={SIAM review},
  volume={45},
  number={2},
  pages={167--256},
  year={2003},
  publisher={SIAM}
}

@article{perry2012null,
  title={Null models for network data},
  author={Perry, Patrick O and Wolfe, Patrick J},
  journal={arXiv preprint arXiv:1201.5871},
  year={2012}
}

@article{zhao2012consistency,
  title={Consistency of community detection in networks under degree-corrected stochastic block models},
  author={Zhao, Yunpeng and Levina, Elizaveta and Zhu, Ji},
  journal={The Annals of Statistics},
  volume={40},
  number={4},
  pages={2266--2292},
  year={2012},
  publisher={Institute of Mathematical Statistics}
}

@article{stochblock,
  title={Estimation and prediction for stochastic blockstructures},
  author={Nowicki, Krzysztof and Snijders, Tom A B},
  journal={Journal of the American statistical association},
  volume={96},
  number={455},
  pages={1077--1087},
  year={2001},
  publisher={Taylor \& Francis}
}

@article{stochblock_degcorrected,
  title={Stochastic blockmodels and community structure in networks},
  author={Karrer, Brian and Newman, Mark EJ},
  journal={Physical review E},
  volume={83},
  number={1},
  pages={016107},
  year={2011},
  publisher={APS}
}

@book{random_cluster,
  title={The random-cluster model},
  author={Grimmett, Geoffrey R},
  volume={333},
  year={2006},
  publisher={Springer Science \& Business Media}
}

@article{selu,
  title={Self-normalizing neural networks},
  author={Klambauer, G{\"u}nter and Unterthiner, Thomas and Mayr, Andreas and Hochreiter, Sepp},
  journal={Advances in neural information processing systems},
  volume={30},
  year={2017}
}

@article{leiden,
  title={From Louvain to Leiden: guaranteeing well-connected communities},
  author={Traag, Vincent A and Waltman, Ludo and Van Eck, Nees Jan},
  journal={Scientific reports},
  volume={9},
  number={1},
  pages={1--12},
  year={2019},
  publisher={Nature Publishing Group}
}

@article{laplacian,
  title={Laplacian dynamics and multiscale modular structure in networks},
  author={Lambiotte, Renaud and Delvenne, J-C and Barahona, Mauricio},
  journal={arXiv preprint arXiv:0812.1770},
  year={2008}
}

@article{resolutionlimit,
  title={Resolution limit in community detection},
  author={Fortunato, Santo and Barthelemy, Marc},
  journal={Proceedings of the national academy of sciences},
  volume={104},
  number={1},
  pages={36--41},
  year={2007},
  publisher={National Acad Sciences}
}

@article{community_detection,
  title={Community detection in graphs},
  author={Fortunato, Santo},
  journal={Physics reports},
  volume={486},
  number={3-5},
  pages={75--174},
  year={2010},
  publisher={Elsevier}
}

@article{newman2004detecting,
  title={Detecting community structure in networks},
  author={Newman, Mark EJ},
  journal={The European physical journal B},
  volume={38},
  number={2},
  pages={321--330},
  year={2004},
  publisher={Springer}
}

@article{graphcuts1,
  title={Normalized cuts and image segmentation},
  author={Shi, Jianbo and Malik, Jitendra},
  journal={IEEE Transactions on pattern analysis and machine intelligence},
  volume={22},
  number={8},
  pages={888--905},
  year={2000},
  publisher={Ieee}
}

@inproceedings{graphcuts2,
  title={Towards efficient hierarchical designs by ratio cut partitioning},
  author={Wei, Yen-Chuen and Cheng, Chung-Kuan},
  booktitle={1989 IEEE International Conference on Computer-Aided Design. Digest of Technical Papers},
  pages={298--301},
  year={1989},
  organization={IEEE}
}

@article{spectral,
  title={On spectral clustering: Analysis and an algorithm},
  author={Ng, Andrew and Jordan, Michael and Weiss, Yair},
  journal={Advances in neural information processing systems},
  volume={14},
  year={2001}
}

@article{newman2006modularity,
  title={Modularity and community structure in networks},
  author={Newman, Mark EJ},
  journal={Proceedings of the national academy of sciences},
  volume={103},
  number={23},
  pages={8577--8582},
  year={2006},
  publisher={National Acad Sciences}
}

@article{louvain,
  title={Fast unfolding of communities in large networks},
  author={Blondel, Vincent D and Guillaume, Jean-Loup and Lambiotte, Renaud and Lefebvre, Etienne},
  journal={Journal of statistical mechanics: theory and experiment},
  volume={2008},
  number={10},
  pages={P10008},
  year={2008},
  publisher={IOP Publishing}
}

@article{cpm,
  title={Narrow scope for resolution-limit-free community detection},
  author={Traag, Vincent A and Van Dooren, Paul and Nesterov, Yurii},
  journal={Physical Review E},
  volume={84},
  number={1},
  pages={016114},
  year={2011},
  publisher={APS}
}

@article{diffpool,
  title={Hierarchical graph representation learning with differentiable pooling},
  author={Ying, Zhitao and You, Jiaxuan and Morris, Christopher and Ren, Xiang and Hamilton, Will and Leskovec, Jure},
  journal={Advances in neural information processing systems},
  volume={31},
  year={2018}
}

@article{agc,
  title={Attributed graph clustering via adaptive graph convolution},
  author={Zhang, Xiaotong and Liu, Han and Li, Qimai and Wu, Xiao-Ming},
  journal={arXiv preprint arXiv:1906.01210},
  year={2019}
}

@article{daegc,
  title={Attributed graph clustering: A deep attentional embedding approach},
  author={Wang, Chun and Pan, Shirui and Hu, Ruiqi and Long, Guodong and Jiang, Jing and Zhang, Chengqi},
  journal={arXiv preprint arXiv:1906.06532},
  year={2019}
}

@inproceedings{topk,
  title={Graph u-nets},
  author={Gao, Hongyang and Ji, Shuiwang},
  booktitle={international conference on machine learning},
  pages={2083--2092},
  year={2019},
  organization={PMLR}
}

@article{dgi,
  title={Deep Graph Infomax.},
  author={Velickovic, Petar and Fedus, William and Hamilton, William L and Li{\`o}, Pietro and Bengio, Yoshua and Hjelm, R Devon},
  journal={ICLR (Poster)},
  volume={2},
  number={3},
  pages={4},
  year={2019}
}

@article{gc_neurosc,
  title={Motifs in brain networks},
  author={Sporns, Olaf and K{\"o}tter, Rolf and Friston, Karl J},
  journal={PLoS biology},
  volume={2},
  number={11},
  pages={e369},
  year={2004},
  publisher={Public Library of Science San Francisco, USA}
}

@article{gc_bio,
  title={Biological network comparison using graphlet degree distribution},
  author={Pr{\v{z}}ulj, Nata{\v{s}}a},
  journal={Bioinformatics},
  volume={23},
  number={2},
  pages={e177--e183},
  year={2007},
  publisher={Oxford University Press}
}

@article{gc_soc,
  title={The strength of weak ties},
  author={Granovetter, Mark S},
  journal={American journal of sociology},
  volume={78},
  number={6},
  pages={1360--1380},
  year={1973},
  publisher={University of Chicago Press}
}

@article{potts_reslimfree,
  title={Local resolution-limit-free Potts model for community detection},
  author={Ronhovde, Peter and Nussinov, Zohar},
  journal={Physical Review E},
  volume={81},
  number={4},
  pages={046114},
  year={2010},
  publisher={APS}
}

@article{potts_temp,
  title={On potts model clustering, kernel k-means and density estimation},
  author={Murua, Alejandro and Stanberry, Larissa and Stuetzle, Werner},
  journal={Journal of Computational and Graphical Statistics},
  volume={17},
  number={3},
  pages={629--658},
  year={2008},
  publisher={Taylor \& Francis}
}

@article{nocd,
  title={Overlapping community detection with graph neural networks},
  author={Shchur, Oleksandr and G{\"u}nnemann, Stephan},
  journal={arXiv preprint arXiv:1909.12201},
  year={2019}
}

@article{gnns_msgpassng,
  title={Relational inductive biases, deep learning, and graph networks},
  author={Battaglia, Peter W and Hamrick, Jessica B and Bapst, Victor and Sanchez-Gonzalez, Alvaro and Zambaldi, Vinicius and Malinowski, Mateusz and Tacchetti, Andrea and Raposo, David and Santoro, Adam and Faulkner, Ryan and others},
  journal={arXiv preprint arXiv:1806.01261},
  year={2018}
}

@article{plantoid,
  title={Collective classification in network data},
  author={Sen, Prithviraj and Namata, Galileo and Bilgic, Mustafa and Getoor, Lise and Galligher, Brian and Eliassi-Rad, Tina},
  journal={AI magazine},
  volume={29},
  number={3},
  pages={93--93},
  year={2008}
}

@article{coauthor,
  title={Pitfalls of graph neural network evaluation},
  author={Shchur, Oleksandr and Mumme, Maximilian and Bojchevski, Aleksandar and G{\"u}nnemann, Stephan},
  journal={arXiv preprint arXiv:1811.05868},
  year={2018}
}

@inproceedings{amazon,
  title={Image-based recommendations on styles and substitutes},
  author={McAuley, Julian and Targett, Christopher and Shi, Qinfeng and Van Den Hengel, Anton},
  booktitle={Proceedings of the 38th international ACM SIGIR conference on research and development in information retrieval},
  pages={43--52},
  year={2015}
}

@inproceedings{microsoft,
  title={An overview of microsoft academic service (mas) and applications},
  author={Sinha, Arnab and Shen, Zhihong and Song, Yang and Ma, Hao and Eide, Darrin and Hsu, Bo-June and Wang, Kuansan},
  booktitle={Proceedings of the 24th international conference on world wide web},
  pages={243--246},
  year={2015}
}

@inproceedings{potts1952,
  title={Some generalized order-disorder transformations},
  author={Potts, Renfrey Burnard},
  booktitle={Mathematical proceedings of the cambridge philosophical society},
  volume={48},
  number={1},
  pages={106--109},
  year={1952},
  organization={Cambridge University Press}
}

@article{potts1990,
  title={On Hammersley’s method for one-dimensional covering problems},
  author={Domb, Cyril},
  journal={Disorder in physical systems},
  pages={33},
  year={1990},
  publisher={Citeseer}
}

@inproceedings{conductance,
  title={Defining and evaluating network communities based on ground-truth},
  author={Yang, Jaewon and Leskovec, Jure},
  booktitle={Proceedings of the ACM SIGKDD Workshop on Mining Data Semantics},
  pages={1--8},
  year={2012}
}

@article{nmi,
  title={Detecting the overlapping and hierarchical community structure in complex networks},
  author={Lancichinetti, Andrea and Fortunato, Santo and Kert{\'e}sz, J{\'a}nos},
  journal={New journal of physics},
  volume={11},
  number={3},
  pages={033015},
  year={2009},
  publisher={IOP Publishing}
}

\end{document}